\documentclass[%
 aip,
 amsmath,amssymb,
 reprint,%
]{revtex4-1}
\usepackage{algorithm}
\usepackage{algorithmicx}
\usepackage{algpseudocode}
\usepackage{color,xcolor}

\usepackage{graphicx}
\usepackage{epstopdf}
\usepackage{dcolumn}
\usepackage{bm}

\usepackage[utf8]{inputenc}
\usepackage[T1]{fontenc}
\usepackage{mathptmx}

\def\BibTeX{{\rm B\kern-.05em{\sc i\kern-.025em b}\kern-.08emT\kern-.1667em\lower.7ex\hbox{E}\kern-.125emX}}

\begin{document}

\preprint{AIP/123-QED}

\title[similarity based on ego-networks]{Measuring similarity in co-occurrence data using ego-networks}

\author{Xiaomeng Wang}
 \email{wxm1706@swu.edu.cn.}
 \altaffiliation{College of Computer and Information Science, Southwest University, Beibei, Chongqing, 400715 P. R. China}
\author{Yijun Ran}%
\affiliation{College of Computer and Information Science, Southwest University, Beibei, Chongqing, 400715 P. R. China
}%
\author{Tao Jia}
\email{tjia@swu.edu.cn.}
\affiliation{College of Computer and Information Science, Southwest University, Beibei, Chongqing, 400715 P. R. China
}%

\date{\today}

%
\begin{abstract}
The co-occurrence association is widely observed in many empirical data. Mining the information in co-occurrence data is essential for advancing our understanding of systems such as social networks, ecosystem, and brain network. Measuring similarity of entities is one of the important tasks, which can usually be achieved using a network-based approach. Here we show that traditional methods based on the aggregated network can bring unwanted in-directed relationship. To cope with this issue, we propose a similarity measure based on the ego network of each entity, which effectively considers the change of an entity's centrality from one ego network to another. The index proposed is easy to calculate and has a clear physical meaning. Using two different data sets, we compare the new index with other existing ones. We find that the new index outperforms the traditional network-based similarity measures, and it can sometimes surpass the embedding method. In the meanwhile, the measure by the new index is weakly correlated with those by other methods, hence providing a different dimension to quantify similarities in co-occurrence data. Altogether, our work makes an extension in the network-based similarity measure and can be potentially applied in several related tasks.
\end{abstract}

\maketitle

\begin{quotation}
The co-occurrence data refer to the type of data where multiple entities simultaneously occur in a single instance, such as the co-tags in folksonomy, the co-author of a scientific paper, co-activation of brain regions under a stimulus, and more. Measuring similarity between entities is fundamental to analyze co-occurrence data, allowing us to further explore social, brain or scientific systems. Using the ego network composed by the co-occurrence relationships as the backbone, we proposed a network-based similarity measure. The new approach outperforms traditional ones and can sometimes surpass the machine learning based embedding method, providing a good tool for tasks such as community detection, link prediction, recommendation.
\end{quotation}

\section{\label{sec:level1}Introduction}
Many tasks in computer science, such as knowledge management \cite{mantymaki2016enterprise, kane2017evolutionary}, community detection \cite{fortunato2016community, dayan2017knowledge}, nature language processing\cite{kumar2016ask, goldberg2016primer} and link prediction \cite{barzel2013network, lu2015toward}, require the measure of similarity between two entities. This can be achieved via different methods based on the nature of the problem analyzed. The similarity would be most straightforward to calculate if the features of the two entities are already mapped into a high dimensional space. Nevertheless, the embedding itself is usually a hard problem and in many cases without a clear physical explanation. Hence, other methods that do not directly use feature vectors are also widely used because of their simplicity and interpretability. For example, if two entities can be expressed by a string, their similarity can be quantified by the minimum number of operations required to transform one string into the other\cite{Navarro2001GTA}. And for time series, Dynamic Time Warping (DTW)\cite{Berndt1994Using,Izakian2015Fuzzy} is the most well known technique for evaluating the similarity with respect to their shape information. Among all of them, network (or graph) based approach is commonly adopted \cite{le2018multiperspective}. Generally, a network is built in which nodes are the entities and links corresponds to association between entities. Similarity is therefore quantified using the direct connection between two nodes or their indirect relationship with other nodes, giving rise to a series of measures, including index based on path length\cite{wu1994verbs,leacock1998combining}, RSS\cite{chen2012discovering}, common neighbors\cite{zhou2009predicting,Chen2015ASCOS} or information theory\cite{li2014new} based index.

In this paper, we focus on similarity measure of entities in co-occurrence data. The co-occurrence association is widely observed in many empirical data, ranging from the co-concepts in pictures\cite{feng2016semantic, henry2019association} to co-words in corpus\cite{cobo2018co, feng2017improving}, from co-authors in publications\cite{mongeon2016costly, wang2019nonlinear, yu2019academic} and co-actors in movies \cite{dang2016timearcs,moreau2017typicality} to co-tags in folksonomy \cite{uddin2013semantic, hellsten2019automated} and co-mention in online communities \cite{barnett2017world, said2019mining}, from the bio-species observed in the same ecosystem \cite{morueta2016network} to co-activated brain regions under a stimuli \cite{bassett2017network, yan2017network}. The similarity measure is a fundamental step towards the understanding of hidden relationships among the co-occurring entities, driving a series of direct application. For example, with similarity measure, we can perform hierarchical clustering in folksonomy data to group tags with similar meanings/semantics; we can better measure associations between different brain regions with similarity, allowing us to further probe the functionality of the brain; we can use similarity to group scientists into different research groups/communities using their co-authorship relations or use similarity in reference relationships to group scientific papers with similar topics, which can help improving the recommendation performance of search engines. In all, similarity between entities is an important measure to analyze the co-occurrence data.

The network based approach to measure similarity in co-occurrence data is to firstly transfer the co-occurred entities into a clique. The clique is a term defined in graph theory which is a network structure where every node (entity) is connected to all others\cite{kat2018matching, xu2018inferring}. These cliques, composed of multiple entities co-appear in the same sample, pile up and eventually form an aggregated network, providing the basis for which information can be mined using the topology features of the network. We find, however, that the aggregated network may bring indirect association between nodes. This weakens the performance of several traditional measures. To cope with the issue, we consider ego network as the backbone and introduce a new similarity measure that is applicable in the ego network. We test our measure using the co-occurrence data of Stack-Overflow programming terms and PACS (Physics and Astronomy Classification Scheme) codes in Physical Review journals. The comparative analysis of indicators shows that our new measure provides very different information compared with existing ones. Hence, it provides a new dimension in quantification of similarity in co-occurrence data, which can be rather useful in tasks such as collaborative filtering. Despite the calculation simplicity, our measure outperforms those based on the aggregated network in predicting similar terms, and sometimes can be better than the embedding method. This can be best illustrated when we apply our new index to cluster Stack-Overflow programming terms, where our new measure yield a very reasonable clustering of terms.

The rest of this paper is organized as follows: Section 2 introduces some related works including similarity indicators based on aggregated network and similarity measure based on the word2vector method. Section 3 introduces the build of ego-networks and the new similarity measure. In section 4 we compare several different indicators and also apply our new measure to a specific hierarchical clustering task. The results support the effectiveness of the proposed index in similarity relationship discovery. Section 5 is the summary of the research.

\section{Related Work}
\begin{figure}[!htb]
    \centering
    \includegraphics[width=\linewidth]{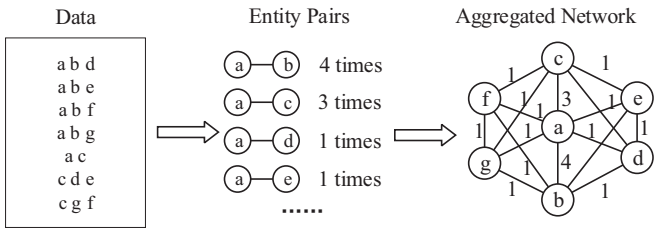}
    \caption{An example of constructing aggregated network using co-occurrence data}
    \label{Fig:ExtractRelationships}
\end{figure}

The co-occurrence usually refers to the instances when two or more entities occur in the same sample\cite{hseu1999image, haralick1973textural}. The network based approach to analyze this kind data is to build a network (Fig.~\ref{Fig:ExtractRelationships}). A node represents an entity and all entities appear in one sample are linked to each other, giving rise to a densely connected subgraph called clique. By combining multiple samples in the data, small cliques are aggregated to a big (aggregated) co-occurrence network, in which the weight of the links correspond to the frequency of two nodes co-occurring. The aggregated network hence provides a structural basis to investigate relationships among nodes.

Among all possible relationships between nodes, the similarity is a simple yet important one intensively studied. The similarity measure in a network can be roughly divided into three categories. The simplest kind just considers the link weight between the two target nodes. The second type, such as relation strength similarity (RSS)\cite{chen2012discovering}, focuses on the relative connection strength by taking the connection strength of both the target nodes and their neighboring nodes into consideration. The relation strength is a normalized edge weight that is calculated as
\begin{equation}
    \sigma_{rss}(i,j) = \frac{\alpha_{ij}}{\sum_{x \in N_i}{\alpha_{ix}}},
\end{equation}
where $\alpha_{ij}$ is the weight of link $(i,j)$ , and $N_i$ is the set of neighbor vertices of node $i$. In order to avoid asymmetric strength, the greater strength value between $\sigma_{rss}(i,j)$ and $\sigma_{rss}(j,i)$ is adopted as follows
\begin{equation}
\sigma_{rss} = \max(\sigma_{rss}(i,j),\sigma_{rss}(j,i)).
\end{equation}

Unlike the former two whose key factor is the link weight, the third type of methods calculate similarity based on the common neighbors of two nodes\cite{Adamic2003Friends}, in the sense that nodes are similar to the extent that their neighborhoods overlap. The simplest index counts the number of common neighbors as
\begin{equation}
\sigma_{cn} = \vert N_i \cap N_j \vert,
\end{equation}
where $N_i$ and $N_j$ are the neighbor sets of node $i$ and $j$, respectively.

Different normalization indexes are also used to quantify the relative strength of common neighbors. Jaccard index\cite{etude1901comparative} normalizes the number of shared nodes based on the neighbor union,
\begin{equation}
    \sigma_{jaccard} = \frac{\vert N_i \cap N_j \vert}{\vert N_i \cup N_j \vert}.
\end{equation}
The cosine similarity proposed by Salton\cite{salton1983mcgill} is defined as
\begin{equation}
    \sigma_{salton} = \frac{\vert N_i \cap N_j \vert}{\sqrt{\vert N_i \vert \vert N_j \vert}}.
\end{equation}
Hub Promoted Index proposed to quantify the topological overlap of pairs of sub-strates in metabolic networks\cite{ravasz2002hierarchical}, is defined as
\begin{equation}
    \sigma_{hpi} = \frac{\vert N_i \cap N_j \vert}{min\{ \vert N_i \vert, \vert N_j \vert \}}.
\end{equation}
Analogously to the above index, Hub Depressed Index\cite{zhou2009predicting} considers the opposite effect on hubs for comparison, which is defined as
\begin{equation}
    \sigma_{hdi} = \frac{\vert N_i \cap N_j \vert}{max\{ \vert N_i \vert, \vert N_j \vert \}}.
\end{equation}

Similarity can also be quantified by machine learning. Representation learning based on deep learning technology can be used to embed entities into a vector space, allowing us to calculate similarity efficiently using closeness of entities in the vector space. The most popular one is word2vec method \cite{mikolov2013efficient,mikolov2013distributed}. The word2vec is successfully applied to measure textual similarity in short context \cite{kenter2015short, shajalal2018sentence} and to make recommendations\cite{wang2018sequence}. Since the number of entities in each sample is usually small in the co-occurrence data, we can use the whole samples (all sets of entities) as the input of word2vec. The similarity between entities is calculated via cosine formula using the vector of each entity that are the output of word2vec.

\begin{algorithm}[H]
\caption{Similarity Calculating}
\label{similaritycalc}
\begin{algorithmic}[1]
\Require The co-occurrence records $R=\{c_1, c_2, ..., c_m\}$ for computing;
\Ensure Similarity Indexes $S=\{\sigma_{ego}(i,j) \vert i \in V, j \in V \}$ where $V$ is the node set;
\State $S \gets \emptyset$;
\State Extract relationships from source data $R$ and create global network $G=(V,E)$;
\State Traverse source data $R$ and extract all triples $T=\{(i,j,k) \vert i \in V, j \in V, k \in V, i \neq j \neq k\}$;
\State Common co-occurrence node set of pair: $PS \gets \emptyset$;
\For{$\forall (i,j,k) \in T$}
\State Add $k$ into $PS[(i,j)]$ which is the co-occurrence node set of $(i,j)$
\State Add $j$ into $PS[(i,k)]$ which is the co-occurrence node set of $(i,k)$
\State Add $i$ into $PS[(j,k)]$ which is the co-occurrence node set of $(j,k)$
\EndFor
\For{$\forall (i,j) \in PS$}
\State $N_i \gets$ the neighbor set of i in $G$
\State $N_j \gets$ the neighbor set of j in $G$
\State $N^i_j \gets$ the number of elements in $PS[(i,j)]$
\State $\sigma_{ego}(i,j) \gets \frac{\vert N^i_j \vert \vert V \vert}{\vert N_i \vert \vert N_j \vert}$
\EndFor\\
\Return  $S$;
\end{algorithmic}
\end{algorithm}

\section{Similarity Based on Ego-Networks}
\begin{figure}[!htb]
    \centering
    \includegraphics[width=\linewidth]{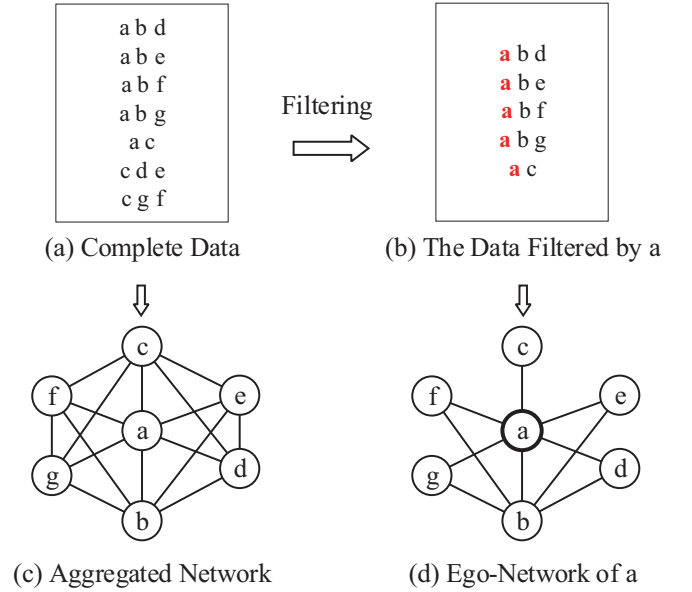}
    \caption{The difference between aggregated network and ego-network. The aggregated network (c) is constructed based on complete data (a) while the ego-network of a (d) is constructed from the data (b) filtered by an entity a.}
    \label{Fig:EgoNetwork}
\end{figure}

The direct and indirect associations are hard to distinguish in the aggregated network. Consequently, two node pairs with different relation strength may have identical neighbors in the aggregated network, yielding inaccurate similarity measures using existing methods. For example, the common neighbors of $(a, b)$ and $(a, c)$ are the same ((Fig.~\ref{Fig:EgoNetwork}(c))) while the relationships can be inferred significantly different based on original records (Fig.~\ref{Fig:EgoNetwork}(a)). In order to cope with this issue, we propose a new method for similarity measure and the idea behind this method is that the direct associations of two objects should dominate the relationship measure while the indirect associations should be eliminated. The direct association between two objects just consider the records where the two objects co-occur. And an ego-network can be used to represent the direct associations between a ``ego'' and its ``alters". As shown in Fig.~\ref{Fig:EgoNetwork}(b) and (d), all records in which object $a$ co-occurs are extracted to construct the ego-network of entity $a$. Then each link to $a$ indicates a direct with between $a$. For instance, $(a, b)$ indicates the co-occurrence of $a$ and $b$, and $(b,g)$ means the co-occurrence of $(a,b)$ and $g$.

To distinguish the aggregated network and ego network, we denote the aggregated network by $G$ and the $i$ centered ego-network by $EG_i$. Because all nodes directly connected with $i$ in $G$ are also connected with $i$ in $EG_i$, $N_i$, which is the neighbor set of $i$ in $G$, is also the neighbor set of $i$ in $EG_i$. However, because the construction of ego network $EG_i$ erases indirect relationships with $i$, the common neighbors between node $i$ and $j$ in $G$ are different from that in $EG_i$. This is illustrated in Fig.~\ref{Fig:EgoNetwork}c and d, where node $a$ and $c$ share 4 common neighbors in $G$ but do not share any common neighbors in $EG_a$. We focus on the neighbor set of node $j$ in $EG_i$, denoted by $N^i_j$. When $i$ and $j$ never co-occur, $N^i_j =0$. When $i$ and $j$ do co-occur but never with other entities, $N^i_j =1$. $N^i_j =N_i$ is the maximum, indicating a strongest relationship such that $j$ appears in all instances when $i$ is located. By normalizing $N^i_j$, we can quantify the relative strength of the coupling between $i$ and $j$ as
\begin{equation}
    C_i(j) = \frac{\vert N^i_j \vert}{\vert N_i \vert}.
\end{equation}

$C_i(j)$ has two limitations. First, it is asymmetric. While $N^i_j=N^j_i$, $N_i$ and $N_j$ are not equal in general, hence $C_i(j) \neq C_j(i)$. Second, $C_i(j)$ does not take any global information into account. Node $i$ and $j$ may be part of a small and isolated cluster, which yields $C_i(j) = 1$, but this strong association is not on the global basis. Therefore, we consider factor $\vert N_j \vert/\vert V \vert$ for correction, where $V$ is the set of all nodes in $G$. This factor quantifies the extend to which node $j$ connects to all nodes, therefore takes both local and global information. Combing $C_i(j)$ and$\vert N_j \vert/\vert V \vert$ together, we finally define the similarity between $i$ and $j$ as
\begin{equation}
    \sigma_{ego}(i,j) = \frac{C_i(j)}{\vert N_j \vert/\vert V \vert} = \frac{\vert N^i_j \vert \vert V\vert}{\vert N_i \vert \vert N_j \vert}.
\end{equation}

$\sigma_{ego}$ is obvious symmetrical. While $\sigma_{ego}$ is derived from the ego network of each entity, we do not necessarily need to construct many ego network to do the calculation. Indeed, the key procedure of $\sigma_{ego}(i,j)$ computing is to count the number of ternary co-occurrences $(i,j,x)$, where $x$ represents a sample in which $(i,j)$ co-occurs. This relationship is readily known when the full data is scanned. Therefore, the complexity for $\sigma_{ego}$ is the same as that for common neighbor based method, and is less complex compared with embedding methods. The concrete steps are described in Algorithm~\ref{similaritycalc}. It is also noteworthy that $\sigma_{ego}$ only considers topological information in the association. For simplicity, link weights are not included. Adding link weights as an extra parameter would definitely improve its performance. As we will demonstrate, even this simple approach would yield results that are as good as ones with link weights considered.

\section{Analysis and Application}
\subsection{Data Description}
The performance of $\sigma_{ego}$ is tested with two data sets: Stack-Overflow data labeled with programming terms (SOFData) and APS publication data labeled with PACS code (PACSData). The details of the two data sets are described as follows.

Stack-Overflow website allows users to label programming terms to different questions. Depending on the content of the question, it may receive multiple tags. For example, ``java'' and ``jvm'' may appear in the same question about java virtual machine. We collect more than 5 million questions from Stack-Overflow web-site, and extract around 45 thousand terms and more than 2 million pairs from the questions. There are multiple types programming terms, such as languages (java, C, etc.), systems (windows, linux, ios, etc.), frameworks (extjs, jquery-mobile, etc.), libraries (pandas, matplotlib, etc.), etc. The similar data set has also been used in other researches and applications\cite{Ponzanelli2014Mining,Chowdhury2015Mining,Abdalkareem2017On}.

PACSData is extracted from publications by American Physical Society (APS) journals. Around 1976, APS started to use PACS codes to label the content of the paper. Each PACS code points to a specific research topic in modern physics and one paper is usually labeled with 1 to 4 PACS codes, carefully chosen by both the author of the paper and the editor of the journal. PACS code can be roughly considered as the key word, but unlike key words which are often created in an ad-hoc, unstructured manner, the PACS code is arranged in a hierarchical format, offering a systematic representation of a paper's subject. Hence, it is wildly used in different studies \cite{jia2017quantifying, zeng2019increasing, battiston2019taking}. The format of the PACS code is ``AB.CD.EF'' which represents a three-level taxonomic relationship. The first two digits 'AB' identify one of 67 top level terms, followed by around 900 second level terms defined by the first four digits 'AB.CD'. There are around 0.3 million articles labeled with PACS codes, giving rise to 48,000 distinct combinations by the first 4 digits of the codes. We use the first two levels of code to compose the network and use the first level to test the similarity prediction. Thanks for the clear classification of PACS code, the accuracy of the similarity prediction can be conveniently checked in PACSData.

The length of sample, which counts the number of entities occur, roughly ranges between 1 to 5. More than half of the samples have the length longer than 3. This naturally illustrates the importance of information mining in co-occurrence data, as most information is hidden in multivariate relationships. Just like most empirical data, the occurrence of the entities combinations follows a power-law distribution: most of them occur only once or twice but there is certain relationship repeatedly occurs in samples.

\subsection{Association Comparison}
\begin{figure}[!htb]
    \centering
    \includegraphics[width=1.0\linewidth]{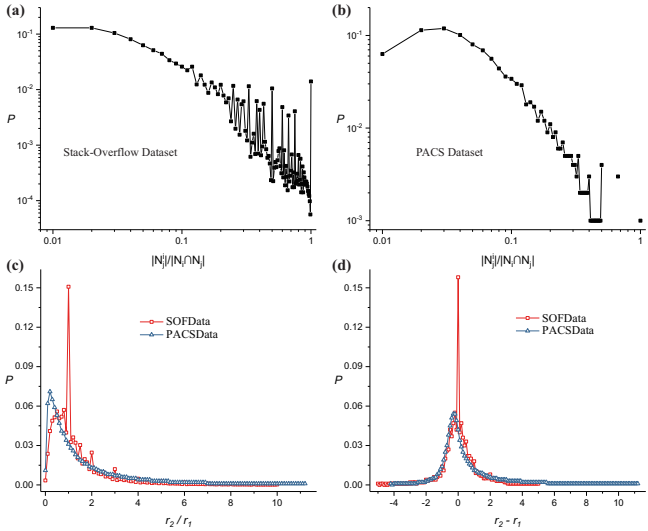}
    \caption{(a, b) The ratio between the number of co-neighbors for $i$ and $j$ in the ego network ($\vert N^i_j \vert$) and in the aggregated network ($\vert N_i\cap N_j \vert$) in two data sets. The distribution decays rapidly, indicating that in almost all cases the common neighbors in the aggregated network and the ego network are different. (c, d) The change in relative relationship between ($x$, $i$) and ($x$, $j$) quantified by ratios ($r_2/r_1$) and differences ($r_2-r_1$). The distribution peaks at 1 in (a) and 0 in (b), indicating that the relative relationship has been reversed when the quantification is switched from $\vert N_i\cap N_j \vert$ to $\vert N^i_j \vert$.}
    \label{Fig:ProbabilityOfNIJ}
\end{figure}

The construction of ego network filters many indirect associations which makes the common neighbor between $i$ and $j$ in the aggregated network ($\vert N_i\cap N_j \vert$) different than that in ego network ($\vert N^i_j \vert$). To give a sense about how big the difference can be, we analyze the ratio $\vert N^i_j \vert /{\vert N_i\cap N_j \vert}$ over our samples. The distributions (Fig.~\ref{Fig:ProbabilityOfNIJ}(a),(b)) indicates that the ratio is small for most pairs, a clear evidence showing how much redundant information is erased by using ego network backbone.

\begin{table*}[htb]
    \caption{Some Samples Whose Associations Change Significantly}
    \label{Tab:contradiction}
    \begin{ruledtabular}
    \begin{tabular}{cccccccc}
    $x$ & $i$ & $j$ & $\vert N_x\cap N_i\vert$ & $\vert N_x\cap N_j\vert$ & $\vert N^x_i\vert$ & $\vert N^x_j\vert$ & $r2 / r1$\\
    \hline
    ...&...&...&...&...&...&...&...\\
    winforms&combobox&python&679&2363&189&11&59.79\\
    latex&pdf-generation&firefox&272&397&36&1&52.54\\
    selenium&rspec&permissions&401&588&67&2&49.12\\
    spring-boot&spring-data-jpa&crash&289&314&131&3&47.44\\
    \textbf{vb.net}&\textbf{textbox}&\textbf{jsp}&834&898&132&3&47.38\\
    c++11&clang&css&660&737&208&7&33.18\\
    mongodb&spring-data&cygwin&321&485&100&7&21.58\\
    ...&...&...&...&...&...&...&...\\
    \textbf{c++builder}&\textbf{cuda}&\textbf{jvcl}&183&27&1&3&0.0492\\
    launch4j&executable&java&45&74&2&67&0.0491\\
    skype&web-services&botframework&222&61&1&23&0.0119\\
    google-docs&caching&google-document-viewer&179&19&1&9&0.0118\\
    ...&...&...&...&...&...&...&...\\
    \end{tabular}
    \end{ruledtabular}
\end{table*}

We further analyze the change of associations between two pairs of entities $(x,i)$ and $(x,j)$. We check the relative change of common neighbors between these two pairs, which determines if $i$ or $j$ is more similar to $x$. Denote $r_1 = \frac{\vert N_x\cap N_i \vert}{\vert N_x\cap N_j \vert}$ and $r_2 = \frac{\vert N^x_i \vert}{\vert N^x_j \vert}$, which measure the relative association strength given by common neighbors, based on aggregated network and ego networks, respectively. If $r_1$ or $r_2$ is greater than 1.0, it means that $i$ may have a stronger association with $x$ than that of $j$. We find that the relative association strength may change from aggregate network to ego networks (Fig.~\ref{Fig:ProbabilityOfNIJ}(c),(d)). We use two indexes $r_2 / r_1$ and $r_2 - r_1$ to measure and analyze such deviation. If $r_2 / r_1 > 1.0$ or $r_2 - r_1 > 0$, it means that the relative association between $x$ and $i$ is enhanced in the ego-network and vice versa. The probability of distribution indicates that most cases, the relative strength changes, implying that the ego network would yield a new measure of similarity different from that in aggregated network.

To give some examples about such difference, we list some example in SOFData whose $r_2 / r_1$ are significantly greater than 1.0 (Table~\ref{Tab:contradiction}). In these samples, term $j$ is predicted to be closer to $x$ than $i$ in the aggregated network, but the ego network predicts the opposite. For example, in aggregated network, ``vb.net'' is closer to ``jsp'' than to ``textbox''. But in reality ``vb.net'' and  ``jsp'' are two unrelated technologies rarely appear in the same question. Their huge common neighbors are solely from the indirect associations. Instead, ``jsp'' should be closer to ``textbox'' because ``textbox'' is a text control component which may be used in ``vb.net'' programming. This relationship is only accurately measured in the ego-network. Likewise,  cuda programs are usually developed by c++ and both terms are popular and generalized concepts, so ``cuda'' and ``c++builder'' have many common neighbors in the aggregated network. On the contrary, ``jvcl'' is a more professional concept, so ``jvcl'' and ``c++builder'' have relatively few co-neighbors in the aggregated network. However, jvcl is a component library for c++ builder and  ``jvcl'' and ``c++builder'' are closer.  This is also reflected in the ego network of ``c++builder''.

\begin{table}
  \caption{Pearson Correlation Coefficient Between Indexes For SOFData}
  \label{Tab:PearsonSOF}
  \begin{ruledtabular}
  \begin{tabular}{cccccccc}
    & $\sigma_{cn}$ & $\sigma_{jaccard}$ & $\sigma_{salton}$ & $\sigma_{hpi}$ & $\sigma_{hdi}$ & $\sigma_{w2v}$ & $\sigma_{ego}$\\
    \hline
    $\sigma_{rss}$ &-0.16 &-0.25 &-0.26 &0.39 &-0.24 &0.09 &0.20\\
    $\sigma_{cn}$ &-&0.50 &0.54 &-0.03 &0.46 &-0.21 &-0.05\\
    $\sigma_{jaccard}$ &-&-&0.94 &-0.46 &0.98 &0.37 &0.12\\
    $\sigma_{salton}$ &-&-&-&-0.27 &0.88 &0.41 &0.11\\
    $\sigma_{hpi}$ &-&-&-&-&-0.52 &-0.04 &0.02\\
    $\sigma_{hdi}$ &-&-&-&-&-&0.33 &0.10\\
    $\sigma_{w2v}$ &-&-&-&-&-&-&0.14\\
  \end{tabular}
  \end{ruledtabular}
\end{table}

\begin{table}
  \caption{Pearson Correlation Coefficient Between Indexes For PACSData}
  \label{Tab:PearsonPACS}
  \begin{ruledtabular}
  \begin{tabular}{cccccccc}
    & $\sigma_{cn}$ & $\sigma_{jaccard}$ & $\sigma_{salton}$ & $\sigma_{hpi}$ & $\sigma_{hdi}$ & $\sigma_{w2v}$ & $\sigma_{ego}$\\
    \hline
    $\sigma_{rss}$ &-0.17 &-0.13 &-0.14 &0.27 &-0.18 &0.33 &0.22\\
    $\sigma_{cn}$ &-&0.76 &0.79 &0.39 &0.73 &-0.21 &-0.06\\
    $\sigma_{jaccard}$ &-&-&0.97 &0.36 &0.97 &0.21 &0.08\\
    $\sigma_{salton}$ &-&-&-&0.51 &0.93 &0.20 &0.06\\
    $\sigma_{hpi}$ &-&-&-&-&0.19 &0.23 &0.07\\
    $\sigma_{hdi}$ &-&-&-&-&-&0.16 &0.06\\
    $\sigma_{w2v}$ &-&-&-&-&-&-&0.16\\
  \end{tabular}
  \end{ruledtabular}
\end{table}
\subsection{Similarity Comparison}
We compare the new index $\sigma_{ego}$ with the seven similarity indicators introduced above. We compute similarity for each pair of entities in the two data sets. Using the similarity value of the same pair but obtained by different indicators, we compute the Pearson correlation coefficient to quantify the similarity between indicators.  The results (Table~\ref{Tab:PearsonSOF}, Table~\ref{Tab:PearsonPACS}) show that $\sigma_{ego}$ is weakly correlated (or approximately independent in some cases) with other index. Note some indicators, such as $\{\sigma_{cn}, \sigma_{jaccard}, \sigma_{salton}, \sigma_{hdi}\}$,  are highly correlated. The average Pearson coefficients among these pairs is 0.72 in SOFData and 0.86 in PACSData. This is a clear evidence that $\sigma_{ego}$ gives a very different dimension of similarity compared with existing ones, which can be very helpful in tasks such as collaborative filtering where complementary criteria are preferred.

\begin{figure}[!htb]
    \centering
    \includegraphics[width=1.0\linewidth]{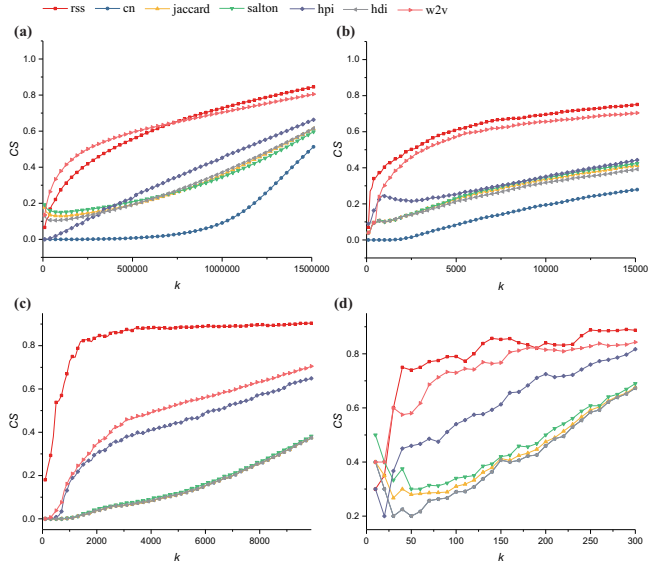}
    \caption{(a, b) The fraction of common pairs in two top-k lists, where one list is fixed and given by $\sigma_{ego}$. The curves indicate that $\sigma_{ego}$ gives very different rank than $\sigma_{cn}$, $\sigma_{jaccard}$, $\sigma_{salton}$, $\sigma_{hpi}$ and $\sigma_{hdi}$. The ranks given by $\sigma_{w2v}$ and $\sigma_{rss}$ are relatively closer to $\sigma_{ego}$.
(c, d) The fraction of common elements in two top-k lists, which predicts the closest terms to ``java'' in (c) and ``42.50'' (quantum optics) in (d). In general, the top-k candidates predicted by $\sigma_{ego}$ is relatively similar to these by $\sigma_{w2v}$ and $\sigma_{rss}$ and different from other indexes.}
    \label{Fig:topcomparation}
\end{figure}

We further check how the rank of pairs sorted by similarity would change. For each similarity index, we rank the pairs of entities in descending order of their similarity values. Use the intersection of two top-k lists, we measure the fraction of common pairs as
\begin{equation}
CS_{ij}^k = \frac{|TK_i\cap TK_j|}{k},
\end{equation}
where $TK_i$ is the Top-k collection based on the indicator $\sigma_{i}$.
The results (Fig.~\ref{Fig:topcomparation}(a),(b)) indicate that the sorting order by $\sigma_{ego}$ is very close to that by $\sigma_{rss}$ and $\sigma_{w2v}$, which both consider the association strength. The sorting order by $\sigma_{ego}$, however, is very different from other common neighbor based index. We also fix an entities and check to what extent the top-k closest terms predicted by different indexes are similar (Fig.~\ref{Fig:topcomparation}(c),(d)). The same conclusion is observed.

\subsection{Predicting Similarity Relationship}
The above discussion confirms that the new index $\sigma_{ego}$ gives very different similarity results than existing ones. It is unclear, however, if it can better detect truly similar entities. To address this point, we compose a positive set (entities that are truly similar) and a negative set (entities that are not similar) for control in SOFData and PACSData. We test the extent that $\sigma_{ego}$ and other indexes can predict similar items against these in the negative set. For SOFData, we rank all term pairs in descending order of their similarity and pick the top 800 pairs. We then artificially judge whether or not two terms have a direct relationship. For example, ``jdbc'' is a database access technology on the ``java'' platform. In other words, ``java'' includes ``jdbc'', hence they are similar. Likewise, because ``mysql'' is an implementation of the ``database'', they are also similar. ``java'' and ``c\#'' are both object-oriented programming (OOP) languages, but they are not directly associated with each other. Therefore, we label them without any direct relationship. We eventually obtain a positive set with 546 pairs and a negative set with 254 pairs. For PACSData, we consider two entities are similar if they share the same first-level code. For example, ``42.25'' and ``42.30'' are similar and ``42.25'' and ``09.11'' are not. We randomly select 1000 pairs from the data that are similar and 1000 pairs that are not similar, composing the positive and negative set, respectively. We use both precision and AUC (the area under the receiver operating characteristic curve) to quantify the prediction performance. For precision, we rank the all pairs in positive and negative set in descending order of their similarity and check the percentage of positive pairs in the top-k list. For AUC, we randomly pick a pair from the positive set and a pair from the negative set, and compare the similarity of the two pairs. If out of $n$ times of independent comparisons, there are $n'$ times that the pair from the positive set has a higher similarity than the pair from the negative set, and $n''$ times that they have the same score, we can calculate $AUC = (n' + 0.5n'')/n$.

\begin{figure}[!htb]
    \centering
    \vspace{-0.2cm}
    \includegraphics[width=1.0\linewidth]{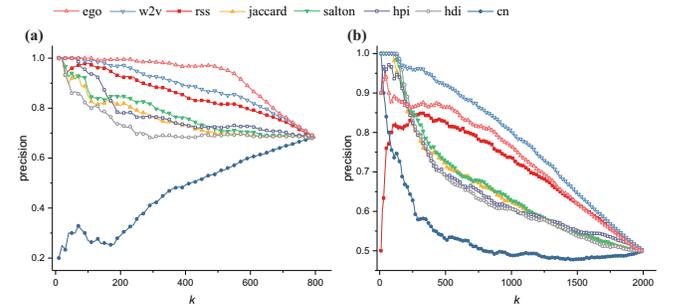}
    \caption{Prediction performance by precision. (a) The result based on the top-k list of 800 term pairs extracted from SOFData (546 positive and 254 negative). (b) The result based on the top-k list of 2000 pairs extracted from PACSData (1000 positive and 1000 negative pairs).}
    \label{Fig:SimilarityPrecision}
\end{figure}

\begin{table}
  \caption{Prediction performance by AUC}
  \label{Tab:AUC}
  \begin{ruledtabular}
  \begin{tabular}{ccccccccc}
    & $\sigma_{rss}$ & $\sigma_{cn}$ & $\sigma_{jaccard}$ & $\sigma_{salton}$ & $\sigma_{hpi}$ & $\sigma_{hdi}$ & $\sigma_{w2v}$ & $\sigma_{ego}$\\
    \hline
    $PACSData$ & 0.798 & 0.477 & 0.520 & 0.531 & 0.506 & 0.505 & 0.761 & {\bf 0.828}\\
    $SOFData$ & 0.784 & 0.193 & 0.577 &0.611 &0.612 & 0.531 & 0.849 & {\bf 0.953}\\
  \end{tabular}
  \end{ruledtabular}
 \label{Table:AUC}
\end{table}

\begin{figure*}[!htb]
    \centering
    \includegraphics[width=1\linewidth]{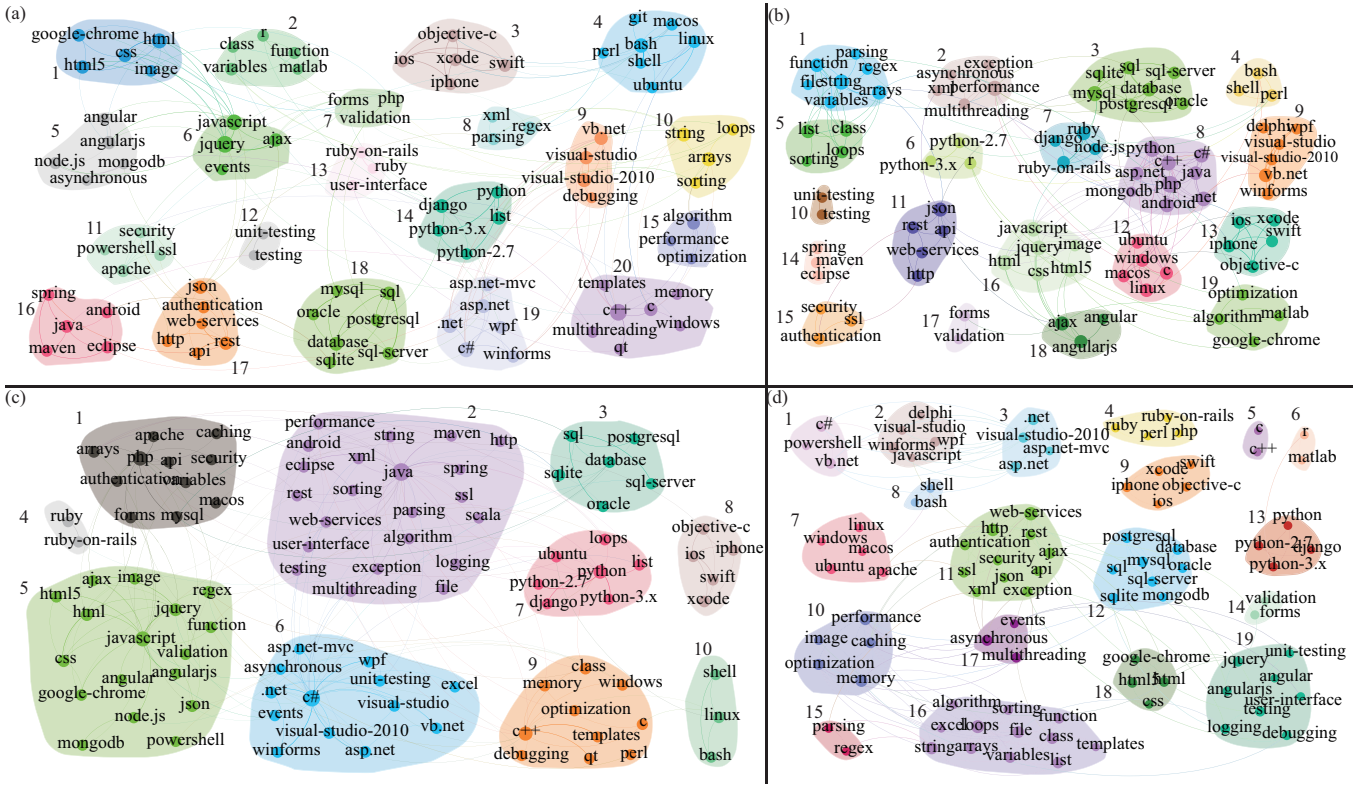}
    \caption{The clustering of the 100 representative terms with the highest degrees in SOFData. Results based on four similarity indicators are listed, (a) by $\sigma_{ego}$, (b) by $\sigma_{jaccard}$, (c) by $\sigma_{rss}$ and (d) by $\sigma_{w2v}$. Different color corresponds to different groups. Isolated nodes, the ones that are not included in any groups, are not shown.}
    \label{Fig:CNNClustring}
\end{figure*}

We find that on the basis of precision, $\sigma_{ego}$ has an obvious advantage compared with other network based indicators using aggregated network (Fig.~\ref{Fig:SimilarityPrecision}). While $\sigma_{ego}$ only considers the topological feature, leaving the association strength (or equivalently link weights) untouched, the precision is already better than link weight based $\sigma_{rss}$. Without using sophisticated embedding methods, the precision of $\sigma_{ego}$ is comparable with word2vector. In SOFData, $\sigma_{ego}$ even is slightly better than word2vector. Given the simplicity in calculation and interpretability with the network based approach, $\sigma_{ego}$ definitely shows some advantages. The performance measured by AUC is even more encouraging. $\sigma_{ego}$ outperforms all other indexes (Table \ref{Table:AUC}). The precision and AUC capture different aspects of the prediction performance. Moreover, the proportion of positive and negative samples will influence the precision but can not influence the AUC. Therefore, the performance evaluation based on the two metrics can be different. But in general, both metrics support the conclusion that our new method is outstanding.

\subsection{Application in Term Clustering}
One direct application of similarity measure is the hierarchical clustering, in which two entities or two communities are merged together based on their similarity. To test our new index, we apply it to cluster terms in SOFData. In this clustering task, we select 100 terms with the largest degree and keep the 200 links/associations with highest similarity value. The clustering result based on $\sigma_{ego}$, $\sigma_{jaccard}$, $\sigma_{rss}$ and $\sigma_{w2v}$ are shown in Fig.~\ref{Fig:CNNClustring}. Despite the fact that ``mongodb'' should have been in database related group (group 18 in Fig.~\ref{Fig:CNNClustring}(a)) and ``r'' and ``matlab'' should have been in different group (but they are included in group 2 in Fig.~\ref{Fig:CNNClustring}(a)), $\sigma_{ego}$ provides a very reasonable division of the terms. On the contrary, both results by $\sigma_{jaccard}$ and $\sigma_{rss}$ have some obvious drawbacks. For example, $\sigma_{jaccard}$ assigns most mainstream programming languages and systems into the same group (group 8 in in Fig.~\ref{Fig:CNNClustring}(b)) and some clusters produced by $\sigma_{rss}$ are not precise enough (group 1, 2, 5 in Fig.~\ref{Fig:CNNClustring}(c)), which are unreasonable from a classification point of view. The result by $\sigma_{w2v}$ does not contain any obvious errors, but it also has some issues (Fig.~\ref{Fig:CNNClustring}(d)). For example, important terms such as ``java'' and ``javascript'' are not clustered to any groups (hence they are not included in the figure as an isolated node), some groups such as 16 and 19 contain too many entities, making it difficult to interpret their meanings.

\section{Conclusion}
The co-occurrence data is a type of data structure rather common in nature. Similarity is an important measure to mine information in this kind of data. Traditional graph based approach uses aggregated network. In this work, we demonstrate evidence that the construction of an aggregated network introduces indirect associations, which wakens the weight of the direct association in the similarity calculation. In order to solve this problem, we proposed a similarity measure based on ego-networks which constructed from the data in which only target ego node occurs.
Our new index is easy to calculate and has a clear physical meaning. The similarity relationship predicted by the new index is better than the traditional index applicable in aggregated networks. The performance is even comparable with the embedding method. The application of this new index to cluster terms in computer science again demonstrates a good performance. Finally, the measure by the new index is weakly correlated with those by other methods, hence providing a different dimension to quantify similarities in co-occurrence data.

For simplicity reasons, our index does not take link weights into consideration. The link weights, measuring how frequent two entities are associated, is an important variable. We believe adding this new feature would significantly improve the performance of the new index. It is also important to apply the index to real systems, to uncover some unknown mechanism. These works are left for future investigations.

\section{Acknowledgments}
This research is supported by the Natural Science Foundation of China (No. 61603309), the Fundamental Research Funds for the Central Universities (XDJK2017C026), and the S-Tech Internet Communication Academic Support Plan.

%

\end{document}